\documentclass[11pt,a4paper]{article}
\pdfoutput=1
\usepackage{jheppubnohead}
\usepackage[utf8]{inputenc}
\usepackage{amsmath}
\usepackage{epsfig}
\usepackage{amssymb}

\providecommand{\mivector}[1]{\vec #1 \,}

\providecommand{\mislash}[1]{#1 \mspace{-10.0mu} \slash}

\providecommand{\proarrow}[0]{\rightarrow}

\providecommand{\dif}[0]{\mathrm{d}}

\providecommand{\proname}[2]{#1 \proarrow #2}

\providecommand{\abs}[1]{\left\lvert #1 \right\rvert}

\providecommand{\abss}[1]{\left\lvert #1 \right\rvert^2}

\providecommand{\miim}[1]{{\rm Im} \left[ #1 \right]}

\providecommand{\order}[1]{{\cal O} \left( #1 \right)}

\providecommand{\gmhor}[2]{\Gamma \left(\proname{#1}{#2}\right)}

\providecommand{\dgmhor}[2]{\Delta \Gamma \left(\proname{#1}{#2}\right)}

\newcommand{\be}{\begin{equation}}
\newcommand{\ee}{\end{equation}}
\newcommand{\bea}{\begin{eqnarray}}
\newcommand{\eea}{\end{eqnarray}}

\title{Unitarity and CP violation in leptogenesis at NLO: general considerations and top Yukawa contributions}
\author[]{J.~Racker}

\affiliation[]{Instituto de Astronom\'{\i}a Te\'orica y Experimental (IATE)\\Universidad Nacional de Córdoba (UNC) - Consejo Nacional de Investigaciones Científicas y Técnicas (CONICET), \\ Laprida 854, X5000BGR,  C\'ordoba, Argentina.\\
Observatorio Astron\'omico de C\'ordoba (OAC), Universidad Nacional de C\'ordoba (UNC),\\ Laprida 854, X5000BGR, C\'ordoba, Argentina. 
}
\emailAdd{jracker@unc.edu.ar}

\abstract{With an emphasis on unitarity and CPT requirements, we study the inclusion of CP-violating processes in baryogenesis at next-to-leading order, particularly  those involving the top Yukawa interaction in leptogenesis. We show that there are more contributions than previously considered, but also important cancellations. Some of these involve the interference of connected with disconnected diagrams. We also discuss on the application of the Kinoshita-Lee-Nauenberg theorem to treat the infrared divergences that are common at next-to-leading order. Finally, we calculate the CP asymmetry in the three-body decay of a sterile neutrino into a lepton and top quarks.
}
\begin{document}

\maketitle
\section{Introduction}

In many baryogenesis scenarios, notably leptogenesis in seesaw models of neutrino masses, the main source of CP violation is typically the out-of-equilibrium decay of a heavy particle, like a sterile neutrino with mass $M$. Here CP violation appears at fourth order in some couplings and it is relatively simple to keep track of all CP-violating processes at that order. However, there are some motivations to consider next-to-leading order (NLO) contributions to the CP-violating source, like those due to the Standard Model (SM) Yukawa or gauge interactions. Indeed, these may actually be dominant at high temperatures ($T\gtrsim M$), while at lower temperatures they can be used to check the precision of  leading order results. Moreover, NLO computations rise interesting theoretical issues.

In standard (type I) leptogenesis, NLO CP-violating processes due to Yukawa and gauge interactions have been incorporated first in~\cite{Pilaftsis:2003gt, Pilaftsis:2005rv, Abada:2006ea} under the hierarchical limit of sterile neutrino masses, which yields CP asymmetries proportional to those in decays. The validity of this hierarchical limit was analyzed in~\cite{Nardi:2007jp, Fong:2010bh} via the explicit calculation of CP-violating processes involving top-Yukawa and gauge interactions. 
Here we show that there are additional, equally important contributions to those formerly considered, some of which lead to important cancellations that are a consequence of unitarity and CPT invariance. Notably, interferences between connected and disconnected diagrams are crucial in this regard. 
 
Special attention to unitarity conditions on CP-violating scatterings was given in~\cite{baldes14} in a neutron portal baryogenesis scenario (see also~\cite{Baldes:2014gca, Baldes:2015lka}). In particular, the authors find that the generation of an asymmetry {\it from scatterings} with only a single out-of-equilibrium heavy particle is not possible. Here we also analyze the validity of this condition in general.

Furthermore, it is interesting to mention the approach that has been taken in~\cite{Bodeker:2017deo}, where a relation between the CP-violating rates and finite-temperature real-time correlation functions was derived. This allowed the authors to obtain explicit expressions at NLO of the CP-violating rates in the hierarchical limit of sterile neutrino masses. Here we calculate the in-vacuum CP asymmetry in the three-body decay of a singlet neutrino at $\order{\lambda_t^2}$, for arbitrary values of the sterile neutrino masses (with $\lambda_t$ being the top Yukawa coupling). However, it is out of the goal of this paper to perform the full computation of all CP-violating rates at $\order{\lambda_t^2}$, and therefore we cannot verify their result. This remains an interesting check for future work. Also notice that leading thermal corrections to the CP asymmetry in Majorana neutrino decays, at first order in the SM couplings, were computed in~\cite{Biondini:2015gyw} and~\cite{Biondini:2016arl}, in the limit of nearly degenerate and hierarchical singlet neutrino masses, respectively. However, in those works the asymmetries at zero temperature were calculated at zeroth order in the SM couplings.

Infrared divergences, which are common in NLO calculations, have typically been cured with thermal masses. However, as pointed out in~\cite{Salvio:2011sf} (see also~\cite{Anisimov:2010gy, Laine:2011pq, Besak:2012qm, Garbrecht:2013gd, Laine:2013lka, Biondini:2013xua, Bodeker:2014hqa, Ghisoiu:2014ena, Bodeker:2015zda} and the review~\cite{Biondini:2017rpb}), these divergences cancel in a more fundamental way by including all processes at a given order in the couplings, as demanded by the Kinoshita-Lee-Nauenberg (KLN) theorem~\cite{Kinoshita:1962ur, Lee:1964is}, which also has its roots in unitarity. Hence, calculations in this work will be done following this approach. 

Actually, as explained in detail in~\cite{Beneke:2014gla}, the full treatment of infrared divergences in the Boltzmann equations (BE) is more involved than the one of vacuum cross sections, where cancellations are guaranteed by the KLN theorem. Indeed, the squared matrix elements of virtual and real corrections to a given process are multiplied by different sets of distribution functions in the BE, given that real corrections involve additional soft and collinear particles. Moreover, the distribution function of bosons diverges as $1/E$ for small momenta, bringing a more severe infrared divergence than the logarithmic ones at zero temperature. As also shown in~\cite{Beneke:2014gla}, it is possible to group the NLO corrections to the BE into a temperature-dependent and a temperature-independent part. This last one involves only vacuum S-matrix elements and it is the only one we will consider in this work.

The paper is organized as follows. In Sec.~\ref{sec:gen} we study some general conditions on baryogenesis derived from unitarity and CPT symmetry. Using Cutkosky rules we show in Sec.~\ref{sec:cut} how, given a certain contribution to a CP-violating process, find another one with exactly the opposite value, so that the conditions found in Sec.~\ref{sec:gen} are satisfied. The CP asymmetry in the three-body decay of a sterile neutrino into a lepton and top quarks is calculated in Sec.~\ref{sec:3bd}, together with some discussions on infrared divergences. Finally,  in Sec.~\ref{sec:con} we summarize the main results and comment on possible directions for further analysis. 

\section{General requirements from unitarity}
\label{sec:gen}

A widely used and simple approximation to follow the evolution of lepton and baryon asymmetries is to set classical BE in an expanding universe, with quantum effects entering only in the calculation of cross sections and decay rates. Moreover, when CP violation is small, an additional good approximation is to linearize the transport equations in the CP-violating quantities. This leads to equations with two types of terms: the so called “washout” terms, proportional to number density asymmetries (reflecting the tendency of the system to approach the equilibrium situation of vanishing asymmetries), and the “source” terms, proportional to the CP asymmetries per scattering or decay process. 

Here we are only interested on outlining some general unitarity-based conditions that apply to the source terms. Although the most basic conditions are well known (see e.g.~\cite{Weinberg:1979bt, Kolb:1979qa}), we make an additional remark and show, in this and the following section, how to apply it in connection with previous works including NLO corrections to the source terms. For simplicity we will neglect quantum statistical effects in the transport equations, hence the resulting equilibrium distribution functions will follow the classical Maxwell-Boltzmann law. The conclusions can be generalized following, e.g., the discussions in~\cite{Weinberg:1979bt, Kolb:1979qa}.

Let us start by writing the BE for the distribution function $f_a=f_a(p,t)$ of a particle whose identity and polarization are collectively denoted by ``$a$'' (e.g., $a$ might be a SM lepton with a definite helicity), and the momentum $p$  is specified in parentheses if necessary to avoid confusion. Given the subtleties that are explained next, it is more convenient to consider the BE for $f_a$ before integrating over the momentum of $a$, namely 

\begin{equation}
\label{eq:befa}
\hat L[f_a] = \sum_{X,Y} f_Y \abss{A(\proname{Y}{a\; X})} - f_a f_X \abss{A(\proname{a\; X}{Y})} \; .
 \end{equation}
 For clarity we have simplified the notation: $\hat L$ denotes the Liouville operator, which in a Robertson-Walker metric is given by
 \begin{equation*}
 \hat L[f_a]= E \frac{\partial f_a}{\partial t} - H(t) \abss{\mivector{p}} \frac{\partial f_a}{\partial E} \;,
 \end{equation*}
with $H(t)$ being the Hubble rate as a function of time and $E$ the energy of the particle. Moreover, $A(\proname{i}{j})$ is the amplitude for the transition $\proname{i}{j}$, $X$ and $Y$ denote sets with an arbitrary number of particles, $f_{X (Y)} \equiv \prod_{x \in X (y \in Y)} f_{x (y)}$, and $\sum_{X,Y}$ is the sum over all possible sets of particles and corresponding phase space,
\begin{equation*}
\sum_{X,Y} \; \longrightarrow \; \sum_{\rm{all \; particle \; sets}}\frac{1}{2} \int \dif \Pi_{X} \dif \Pi_{Y} (2 \pi)^4 \delta^4(p_a + P_X - P_Y) \; .
\end{equation*} 
Here $P_{X (Y)} \equiv \sum_{x \in X (y \in Y)} p_{x (y)}$, $p_x$ is the momentum of particle $x$, $\dif \Pi_{X (Y)} \equiv \prod_{x \in X (y \in Y)} \dif \pi_{x (y)}$ and
\begin{equation*}
\dif \pi_x \equiv \frac{g_x}{(2 \pi)^3} \frac{\dif^3p_x}{2 E_x}  \; ,
\end{equation*}
with $g_x$ the number of internal degrees of freedom of $x$. 

An important consideration is whether a process like $X a(p) \to Y a(p)$, that leaves the number of ``$a$'' particles and their momentum invariant, should be included in Eq.~\ref{eq:befa}. It might seem that this issue is irrelevant, given that the contribution of such process should be negligible when integrating over all possible momenta. However, it turns out that there are in general finite contributions to $X a(p) \to Y a(p)$ coming from the interference of connected with disconnected diagrams, as those depicted in Fig.~\ref{fig:scat}. Then one may choose between, (i) explicitly exclude $X a(p) \to Y a(p)$ from the BE for $f_a(p,t)$, or (ii) include it both in the production and destruction terms, with opposite signs, so that the overall contribution is null. Of course (i) and (ii) are equivalent, but (ii) makes it neater to apply the unitarity conditions below, which involve a sum over all possible initial or final states. Therefore we will use the approach (ii) in this work. Moreover, for our purposes it is enough to take ``$a$'' to be a fermion. Hence, the amplitude of processes involving two or more $a(p)$ in the initial or final state is zero, and therefore they do not contribute to the BE or the unitarity conditions.

Similarly to Eq.~\ref{eq:befa}, the BE for the distribution function of the antiparticle $\bar a$ is
\begin{equation}
\label{eq:befba}
\hat L[f_{\bar a}] = \sum_{X,Y} f_{\bar{Y}} \abss{A(\proname{\bar{Y}}{\bar{a}\; \bar{X}})} - f_{\bar a} f_{\bar X}  \abss{A(\proname{\bar{a}\; \bar{X}}{\bar{Y}})}\; .
\end{equation}
We are interested in possible differences in the evolution of $f_a$ and $f_{\bar a}$, therefore we subtract Eq.~\ref{eq:befba} to Eq.~\ref{eq:befa}. At lowest (linear) order in the CP-odd quantities $\Delta{\abss{A(\proname{i}{j})}} \equiv \abss{A(\proname{i}{j})} - \abss{A(\proname{\bar i}{\bar j})}$ and $\Delta f_i \equiv f_i - f_{\bar i}$, the subtraction yields
 \begin{equation}
 \begin{split}
\label{eq:bedeltaf}
\hat L[\Delta f_a] = & \sum_{X,Y} f_Y \Delta \abss{A(\proname{Y}{a\; X})} - f_a f_X \Delta \abss{A(\proname{a\; X}{Y})} +\\
& \sum_{X,Y} \Delta f_Y \abss{A(\proname{Y}{a\; X})} - \left( \Delta f_a  + \Delta f_X\right) \abss{A(\proname{a\; X}{Y})} \; .
\end{split}
\end{equation}
The terms in the second line are proportional to $\Delta f_i$ and hence give the washout part of the BE, while the source terms in the first line are of prime importance for our analysis, since no asymmetry  $\Delta f_i$ can be generated when they are zero. 

The source part of the BE has contributions from: (I) -the subtraction of- production processes of $a$ and $\bar a$, (II) -the subtraction of- destruction processes of $a$ and $\bar a$. Unitarity and CPT imply that
\begin{equation}
\label{eq:unia}
\sum_Y \Delta \abss{A(\proname{a\; X}{Y})} = 0 \;.
\end{equation}
Therefore {\it the total contribution of the destruction terms to the source is zero}.

Unitarity combined with  CPT symmetry also implies  
\begin{equation}
\label{eq:unib}
\sum_Y \Delta \abss{A(\proname{Y}{a\; X})} = 0 \;.
\end{equation}
This condition -fortunately- cannot be applied directly to the BE because the probabilities $\abss{A(\proname{Y}{a\; X})}$ enter weighted by the distribution functions of the particles in each set $Y$. However, for a particle in equilibrium,  $f= f^{\rm eq}=e^{-E/T}$~\footnote{There could eventually be other conserved quantities besides the energy, but they would also appear as linear combinations in the exponential and therefore the conclusions would not change.}, therefore if all the particles belonging to all possible sets $Y$ are in equilibrium, $f_Y$ would be the same for all $Y$ -fixing $E_a$ and $E_X$-, namely $f_Y=e^{-(E_a+E_X)/T}$. This fact combined with the condition in Eq.~\ref{eq:unib} would imply a null source term (thus arriving at the Sakharov third condition~\cite{sakharov67}). 

Next assume that there is a particle $N$ belonging to at least one of the sets $Y$,  that is out of equilibrium.
Because of the preceding argument, if we write $f_N=f_N^{\rm eq} + (f_N - f_N^{\rm eq})$ it is clear that 
\begin{equation*}
\sum_{Y'} f_N^{\rm eq} f_{Y'}^{\rm eq} \; \Delta \abss{A(\proname{N\, Y'}{a\; X})} + \sum_{Y''} f_{Y''}^{\rm eq} \; \Delta \abss{A(\proname{Y''}{a\; X})}= 0 \;,
\end{equation*}
where all the particles belonging to $Y', Y''$ are taken to be in equilibrium.
Therefore, the BE at linear order in the CP-odd quantities and with only one species $N$ being out of equilibrium, necessarily has the form
\begin{equation}
\label{eq:bedeltaf}
\hat L[\Delta f_a] = \sum_{X,Y'} (f_N - f_N^{\rm eq}) f_{Y'}^{\rm eq} \, \Delta\abss{A(\proname{N\, Y'}{a\, X})}  - \sum_{X,Y} \left( \Delta f_a  + \Delta f_X - \Delta f_Y \right) \abss{A(\proname{a\, X}{Y})},
\end{equation}
where we have also used the CPT condition $\abss{A(\proname{Y}{a\; X})} = \abss{A(\proname{\bar a\; \bar X}{\bar Y})}$ to simplify the washout term. Note that up to this point the discussion has been very general, with no need to fix the model, specify all the relevant processes and eventually apply real intermediate state subtractions. Furthermore notice that, actually, Eq.~\ref{eq:bedeltaf} is only valid when there is at most one $N$ in each set $Y$, but the generalization is trivial: if $N$ appears $r$ times in a certain set of particles $Y$, the corresponding factor in the BE is $(f_N^r - f_N^{{\rm eq}\, r})$.

Although unitarity and CPT requirements on the BE for baryogenesis are known to be fundamental~\cite{sakharov67} and have been extensively discussed (see, e.g., the classic works~\cite{Weinberg:1979bt, Kolb:1979qa}),  usually the derivation of the source term is not presented as simply as we have done above.
 Notably, a conclusion that stands out is that {\it the only contributions to the source of the BE for $\Delta f_a$ come from production processes of ``$a$'' (or ``$\bar a$'') particles, with an out-of-equilibrium species $N$ in the initial state}.  This statement can be useful to avoid introducing spurious contributions, but must be interpreted with care, since it holds as long as the processes of the kind $Y a(p) \to X a(p)$ are handled according to the approach (ii) described above. That is to say, $Y a(p) \to X a(p)$ is included twice in the BE for $f_a$, once with a negative sign (corresponding to the destruction of an $a(p)$), and another with a positive sign (corresponding to the production of an $a(p)$). Something similar is done with the BE for $f_{\bar a}$. After summing all terms in the BE for $\Delta f_a$, the contributions to the source from all destruction processes, including  $Y a(p) \to X a(p)$, cancel, but $Y a(p) \to X a(p)$ might still appear in the source term, because it also produces an $a(p)$.  
 
 To illustrate this issue, consider the scattering into top quarks, $\proname{N_1 \ell_\alpha}{q \bar t}$, of a sterile neutrino $N_1$ and a SM lepton doublet $\ell_\alpha$ playing the role of ``$a$''. In~\cite{Nardi:2007jp} the CP asymmetry  $\sigma (\proname{N_1 \ell_\alpha}{q \bar t})$ -  $\sigma (\proname{N_1 \bar \ell_\alpha}{\bar q t})$ was calculated (with $\sigma$ designating a cross section), finding in particular two contributions from two different ways of cutting the vertex one-loop diagram, that we denote by $C_1$ and $C_2$ in Fig.~\ref{fig:scat}. The statement in the previous paragraph is useful because it warns that there have to be other CP-violating processes cancelling the contributions from $\proname{N_1 \ell_\alpha}{q \bar t}$ to the source term of the lepton asymmetry. This will be analyzed in the following section, but we anticipate that one of the contributions ($C_2$) is cancelled by another coming from $\proname{N_1 \ell_\alpha}{N_2 \ell_\beta}$ (not included in previous works). However, the contribution from the other cut ($C_1$) cancels with that coming from $\proname{N_1 \ell_\alpha}{\ell_\beta \phi \ell_\alpha}$, when it is interpreted as a destruction process of $\ell_\alpha$ in the BE. Therefore, following the approach (ii),  the contribution from $C_1$ ``reappears'' in  $\sigma (\proname{N_1 \bar \ell_\alpha}{\bar \ell_\beta \bar \phi \bar \ell_\alpha})$ -  $\sigma (\proname{N_1 \ell_\alpha}{\ell_\beta \phi \ell_\alpha})$, given that  $\proname{N_1 \ell_\alpha}{\ell_\beta \phi \ell_\alpha}$ also enters the BE as a production process with an out-of-equilibrium particle ($N_1$) in the initial state. Alternatively, if one follows the approach (i), the process $\proname{N_1 \ell_\alpha}{\ell_\beta \phi \ell_\alpha}$, although entering in the unitarity condition $\sum_f \Delta \abss{A(\proname{N_1 \ell_\alpha}{f})} = 0$, is not included at all in the BE for  $\Delta f_{\ell_\alpha}$. From this point of view, the contribution from $C_1$ ``survives'' in the BE, because it is cancelled in the unitarity condition by a process that does not change the number of $\ell_\alpha$. In any case, the safe procedure is to verify that all processes required to satisfy the unitarity conditions have been taken into account.
 
In Sec.~\ref{sec:cut} we will show how to easily find the contributions that cancel the CP violation coming from, e.g., $\proname{N_1 \ell_\alpha}{q \bar t}$, but before finishing this section we wish to comment on a statement made in~\cite{baldes14}. 
Namely, they find that the generation of an asymmetry from scatterings (as opposed to decays) with only a single out-of-equilibrium heavy particle is not possible. They show this for a neutron portal baryogenesis scenario, but here we argue that this statement is not valid in general. 
From the above considerations it can be seen that it is not necessary that there be two out-of-equilibrium species to have a source term, induced by scatterings, in the BE of a {\it single} particle $a$. Nevertheless, to get the total baryon (or lepton) asymmetry, one is interested in summing $\Delta f_a$ over all particles that carry baryon (or lepton) number. Doing so, Eq.~\ref{eq:bedeltaf} becomes
\begin{equation}
\label{eq:sumbedeltaf}
\hat L\left[\sum_a \Delta f_a\right] =  \sum_{X,Y',a} (f_N - f_N^{\rm eq}) f_{Y'}^{\rm eq}  \, \Delta \abss{A(\proname{N\, Y'}{a\, X})}  - \; {\rm washout \; terms},
\end{equation}
where the sum over $a$ runs over all the particles with positive baryon (or lepton) number. Unitarity and CPT symmetry imply that
\begin{equation}
\label{eq:unic}
\sum_f \Delta \abss{A(\proname{N\, Y'}{f})} = 0 \;.
\end{equation}
Therefore, it seems that performing first the sum over $X$ and $a$ in Eq.~\ref{eq:sumbedeltaf} would result in a null source, from which 
the statement in~\cite{baldes14} would follow. However, there are two caveats to this argument. First, if the state $N\, Y'$ is self-CP-conjugate (e.g. consisting of two Majorana particles, with only one being out of equilibrium), then the sum over $X$ and $a$ does not cover all possible final states ($\proname{N\, Y'}{\bar a \bar X}$ is also allowed), therefore Eq.~\ref{eq:unic} cannot be applied (this is the same reason why leptogenesis via Majorana neutrino decays is viable). Second, even if the source term in Eq.~\ref{eq:sumbedeltaf} is zero, this does not preclude a baryon asymmetry from being generated. This is, for instance, the case in ``purely flavoured leptogenesis''~\cite{aristizabal09b}, where flavour effects allow a baryon asymmetry to be generated even when the total CP asymmetry (i.e. summed over all lepton flavours) is zero.

\section{Cancellations from cutting rules}
\label{sec:cut}
Consider the unitarity condition in Eq.~\ref{eq:unia} (Eq.~\ref{eq:unib} is just the CPT conjugate), which was used in the derivation of Eq.~\ref{eq:bedeltaf} to show that the total contribution of the destruction processes to the source is zero. We are going to show that for each contribution to a CP asymmetry in Eq.~\ref{eq:unia}, there is another one with the same magnitude but opposite sign, thus cancellations come in pairs. These may be easily found using Cutkosky rules~\cite{Cutkosky:1960sp, Veltman:1994wz}. For other diagrammatic analyses of unitarity in different baryogenesis contexts see~\cite{roulet97, Hook:2011tk}. 

CP violation requires both, a relative CP-even and a relative CP-odd phase. Specifically, writing an amplitude as the sum of two contributions with the couplings factored into the parameters $\lambda_i$, i.e. $A(\proname{i}{j}) = \lambda_0 I_0 + \lambda_1 I_1$, one gets 
\begin{equation}
\label{eq:deltaa1}
\Delta \abss{A(\proname{i}{j})} = - 4 \miim{\lambda^*_0 \lambda_1} \miim{I^*_0 I_1} \; .
\end{equation} 
Taking $I_0$ to be a -real- tree level contribution,
the cutting rules yield
\begin{equation}
\label{eq:deltaa2}
\miim{I^*_0 I_1} = \frac{1}{2i} I_0 \sum_{\rm cuts} I_1 \; ,
\end{equation}
where the sum runs over all possible ways of cutting the diagrams contributing to $I_1$, such that all cut propagators can be put on-shell.  Now consider the contribution of one of these terms to Eq.~\ref{eq:unia}. It consists of the product of some couplings and three additional factors: a tree level factor ($I_0$), the part of $I_1$ coming from the left of the cut ($I_1^{\rm L}$), and the one from the right ($I_1^{\rm R}$). The key point is that these three factors appear in another term contributing to Eq.~\ref{eq:unia}, but with the corresponding couplings conjugated, therefore cancelling the former contribution. This second term arises from the interference of: (i) a tree level diagram equal to the left part of the cut diagram of the first term, and (ii) a cut diagram whose left part is the tree level diagram of the first term and the right part is also the right part of the cut diagram of the former term, but with the arrows reversed. 

To illustrate this, let us resume the discussion on the annihilation of a sterile neutrino and a SM lepton doublet, within the seesaw model, that we started in the previous section. The lagrangian in the mass basis of the singlet neutrinos $N_i$ reads
\begin{equation}
\label{eq:lag}
\mathcal{L} = \mathcal{L_{\rm SM}} + \mathcal{L_{\rm kin}} - \frac{1}{2} M_i \bar N_i^c N_i  - \lambda^*_{\alpha i} \bar \ell_\alpha \tilde \phi N_i + {\rm h.c.} \;,
\end{equation}
where there is an implicit sum over repeated flavour indices, $\ell_\alpha$ are the leptonic $SU(2)$ doublets, $\phi$ is the Higgs field and $\tilde \phi =i\tau_2 \phi^*$, with $\tau_2$ Pauli's second matrix. Furthermore, we denote by $t$ the right-handed top singlet, by $q$ the left-handed quark doublet containing the top quark, and by $\lambda_t$ the top Yukawa coupling. 

\begin{figure}[!t]
\centerline{\protect\hbox{
\epsfig{file=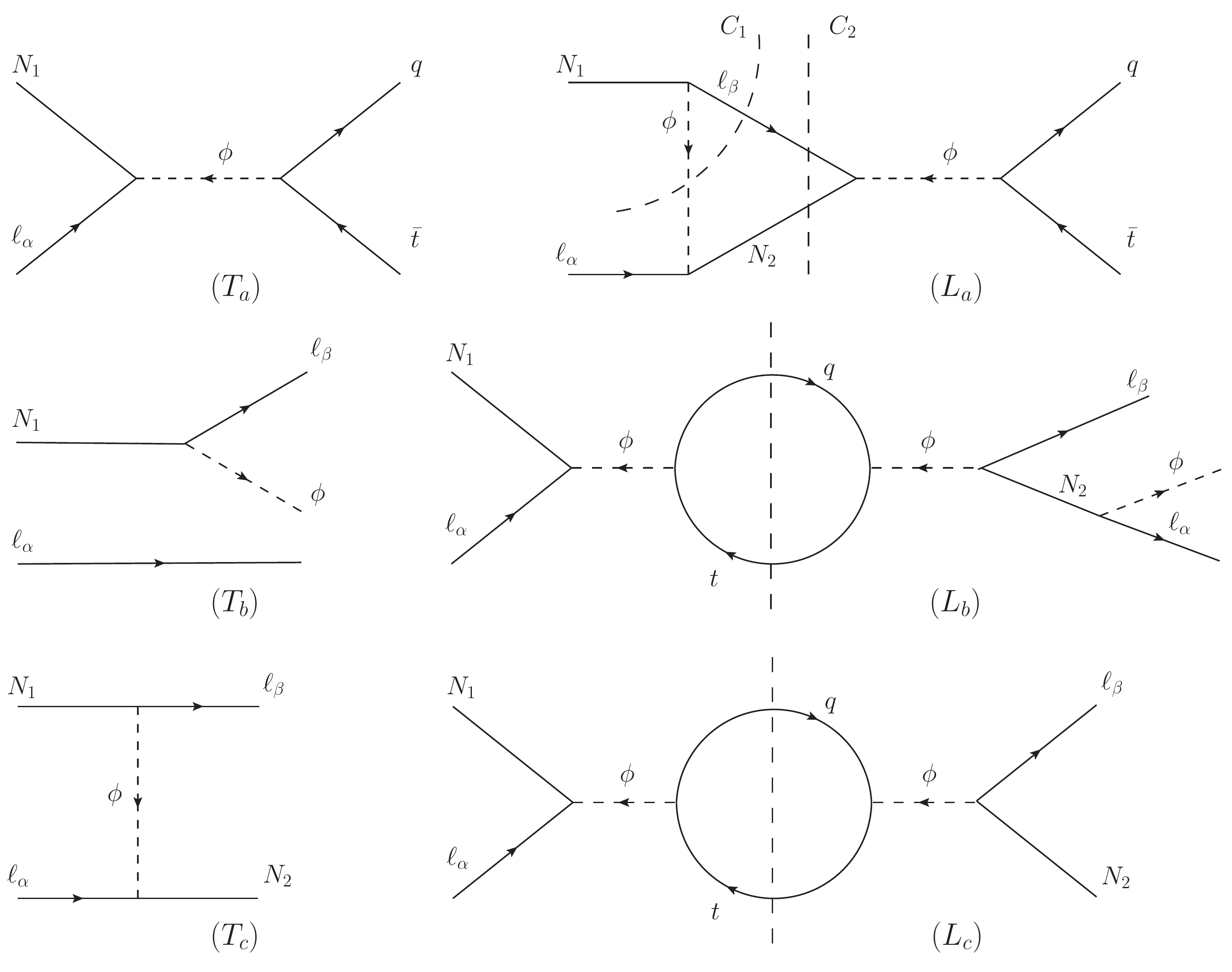,width=0.95\textwidth,angle=0}}}
\caption[]{Some of the diagrams contributing to $\sum_f \Delta \abss{A(\proname{N_1 \ell_\alpha}{f})}$.  Tree-level and one-loop diagrams are denoted with a $T$ and an $L$, respectively. The vertical dashed lines indicate the cuts that yield an imaginary amplitude. For $L_a$ there are two possible cuts that have been denoted by $C_1$ and $C_2$. All Feynman diagrams in this work have been drawn using JaxoDraw~\cite{Binosi:2003yf}.} 
\label{fig:scat}
\end{figure}

The diagrams of four different contributions to the sum $\sum_f \Delta \abss{A(\proname{N_1 \ell_\alpha}{f})}$ are shown in Fig.~\ref{fig:scat}. The interference of the diagrams at the top of the figure gives two contributions to $\Delta \abss{A(\proname{N_1 \ell_\alpha}{q \bar t})}$ because there are two ways of putting the intermediate particles on-shell (the cut $C_2$ requires the center-of-mass energy to be larger than $M_2$). Following the procedure explained above, we have drawn the diagrams in the middle and bottom of the figure. The ones in the middle have been constructed as follows: $T_b$ is the part at the left of the cut $C_1$ of diagram $L_a$, the left part of $L_b$ is $T_a$, and the right part is that part of $L_a$ at the right of cut $C_1$, with the arrows inverted.  The interference of $T_b$ with $L_b$ yields a contribution to $\Delta \abss{A(\proname{N_1 \ell_\alpha}{\ell_\beta \phi \ell_\alpha})}$ equal in magnitude, but opposite in sign, to the contribution to $\Delta \abss{A(\proname{N_1 \ell_\alpha}{q \bar t})}$ coming from the interference of $T_a$ with the cut $C_1$ of $L_a$. Hence both pieces cancel in $\sum_f \Delta \abss{A(\proname{N_1 \ell_\alpha}{f})}$. Likewise, the diagrams at the bottom of Fig.~\ref{fig:scat} come from a reordering of the ones at the top: $T_c$ is the part at the left of cut $C_2$, the left part of $L_c$ is $T_a$, and the right part is given by reversing the arrows in that part of $L_a$ at the right of $C_2$. $T_c$ and $L_c$ interfere to give a term in  $\Delta \abss{A(\proname{N_1 \ell_\alpha}{N_2 \ell_\beta})}$, exactly opposite in value to the contribution of the cut $C_2$ to  $\Delta \abss{A(\proname{N_1 \ell_\alpha}{q \bar t})}$. Therefore, again there is a cancellation in  $\sum_f \Delta \abss{A(\proname{N_1 \ell_\alpha}{f})}$. We have explicitly verified at $\order{\lambda_t^2}$ that all terms in $\sum_f \Delta \abss{A(\proname{N_1 \ell_\alpha}{f})}$ cancel in pairs, in the way we have explained.

\begin{figure}[!t]
\centerline{\protect\hbox{
\epsfig{file=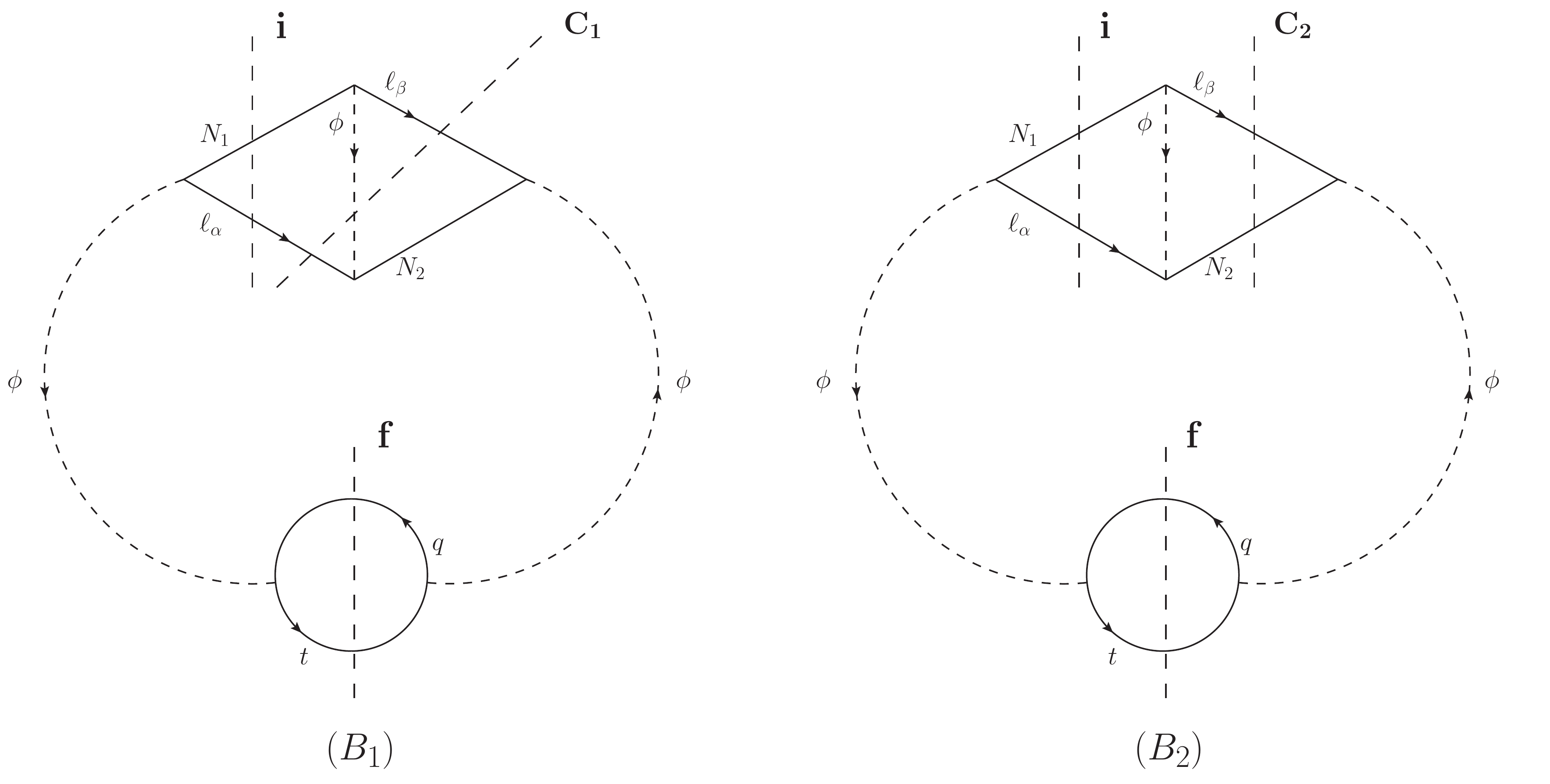,width=0.95\textwidth,angle=0}}}
\caption[]{Closed Feynman diagrams with three cuts, representing the interference of tree-level and one-loop amplitudes contributing to CP violation. The three cuts denoted by $i$, $f$, and $C_i$ determine the initial state, final state, and Cutkosky cut, respectively. The graphs $(B1)$ and $(B_2)$ correspond to the interference of the diagrams $T_a$ and $L_a$ cut along $C1$ and $C2$, respectively. Permuting the cuts $f$ and $C_1$ ($C_2$), yields the interference of the diagrams in the middle (bottom) of Fig.~\ref{fig:scat}. } 
\label{fig:closed}
\end{figure}

Another way to see these cancellations is via closed diagrams with three cuts, one to specify the initial state, another for the final state, and the third one to apply Cutkosky rules, like in Fig.~\ref{fig:closed}. Each choice of cuts determines a contribution to a CP asymmetry, according to Eqs.~\ref{eq:deltaa1} and~\ref{eq:deltaa2}. Similar graphs were used in~\cite{roulet97} to demonstrate diagrammatically that the total CP asymmetry in leptonic scatterings mediated by Majorana neutrinos vanishes at lowest non-trivial order. Here we note that for a given closed diagram, all the permutations of the three cuts yield contributions of the same magnitude to -different- CP violating processes. The relative sign depends on whether the tree level or one loop amplitude is obtained when going clockwise from the initial to the final state cut. For example, keeping the initial-state cut fixed and permuting the other two cuts, gives pairs of cancelling contributions to Eq.~\ref{eq:unia}. 

The processes in the middle and at the bottom of Fig.~\ref{fig:scat} were not considered in previous works on leptogenesis including CP-violating scatterings. The contribution of the cut $C_2$ to the source term found in~\cite{Nardi:2007jp} is cancelled by another contribution coming from $\proname{N_1 \ell_\alpha}{N_2 \ell_\beta}$. However, the contribution from the cut $C_1$ remains in the source term, as discussed in the previous section. Moreover, relabelling flavour indices in Fig.~\ref{fig:scat}, new contributions to the BE for $\Delta f_{\ell_\alpha}$  become apparent, like $\proname{N_1 \ell_\beta}{\ell_\beta, \phi, \ell_\alpha}$.  Especially worth noticing are possible interferences between connected and disconnected diagrams, like $L_b$ with $T_b$. Disconnected diagrams have also been found to be necessary to cancel infrared divergences, as demanded by unitarity~\cite{Lee:1964is} (see also~\cite{Lavelle:2005bt, Lavelle:2010hq, Frye:2018xjj}). This point will be discussed in somewhat more detail in the next section.

\section{CP asymmetry in three-body decays}
\label{sec:3bd}
In order to further illustrate the issues discussed above, it is interesting to consider the CP asymmetry in the three-body decay $\proname{N_i}{\ell_\alpha \bar q t}$. Another motivation is that this is expected to give the dominant contribution to CP violation at $\order{\lambda_t^2}$ when decays start to dominate over scatterings at $T \lesssim M_i$ (see e.g.~\cite{giudice04}).

\begin{figure}[!t]
\centerline{\protect\hbox{
\epsfig{file=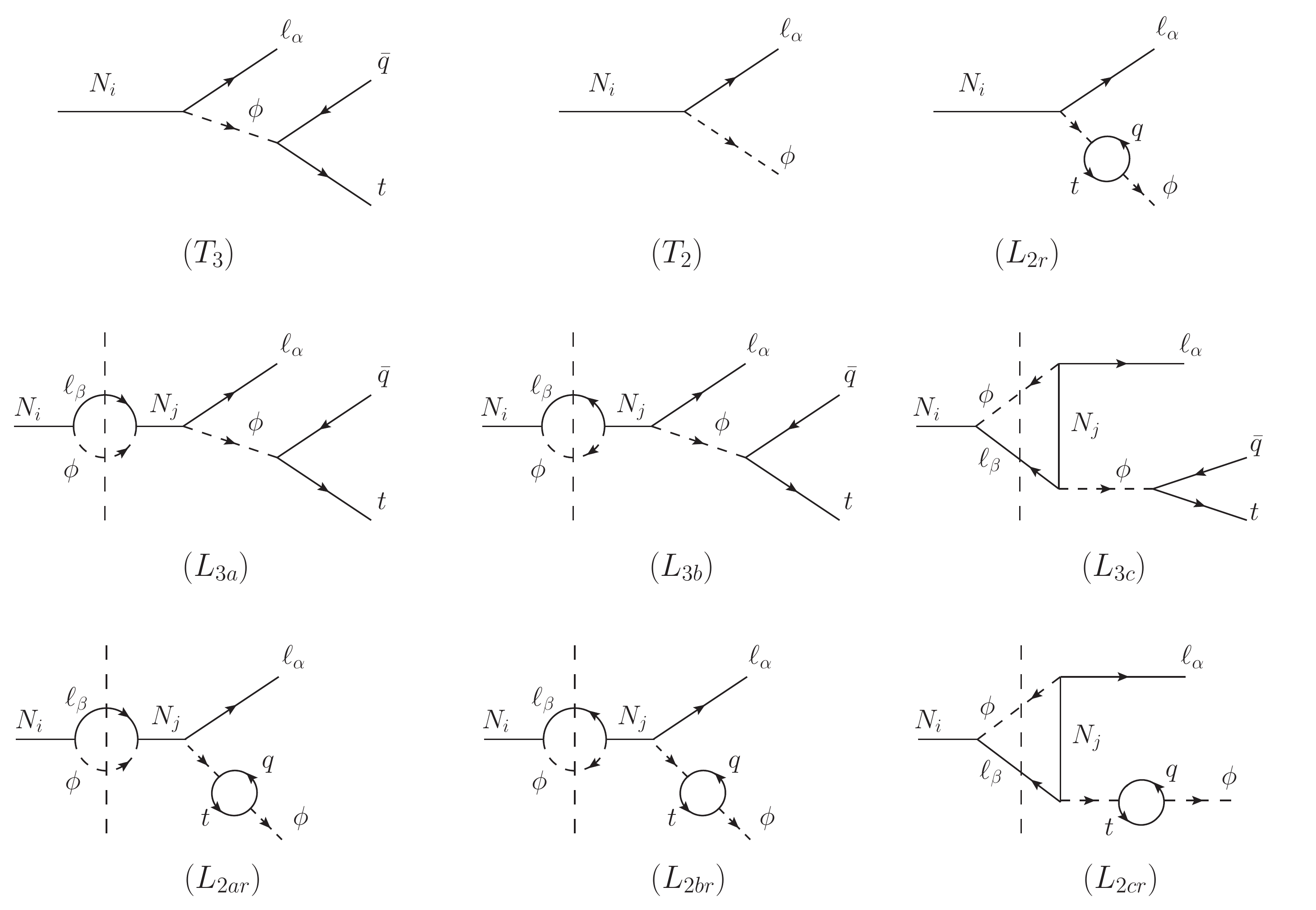,width=0.95\textwidth,angle=0}}}
\caption[]{Some of the diagrams contributing to the CP asymmetry in two- and three-body decays at $\order{\lambda_t^2}$. The vertical dashed lines indicate the possible cuts to obtain an imaginary amplitude. We have denoted with a $T$, the tree level diagrams, and with an $L$, the ones with loops. Those with the top quark loop in a Higgs line carry an additional $r$ subscript. The complete list of diagrams also includes the ones similar to those at the bottom but without the quark loops, to be denoted by $L_{2a, 2b, 2c}$, and the diagrams in Fig.~\ref{fig:2bd}.} 
\label{fig:3bd}
\end{figure}

\begin{figure}[!t]
\centerline{\protect\hbox{
\epsfig{file=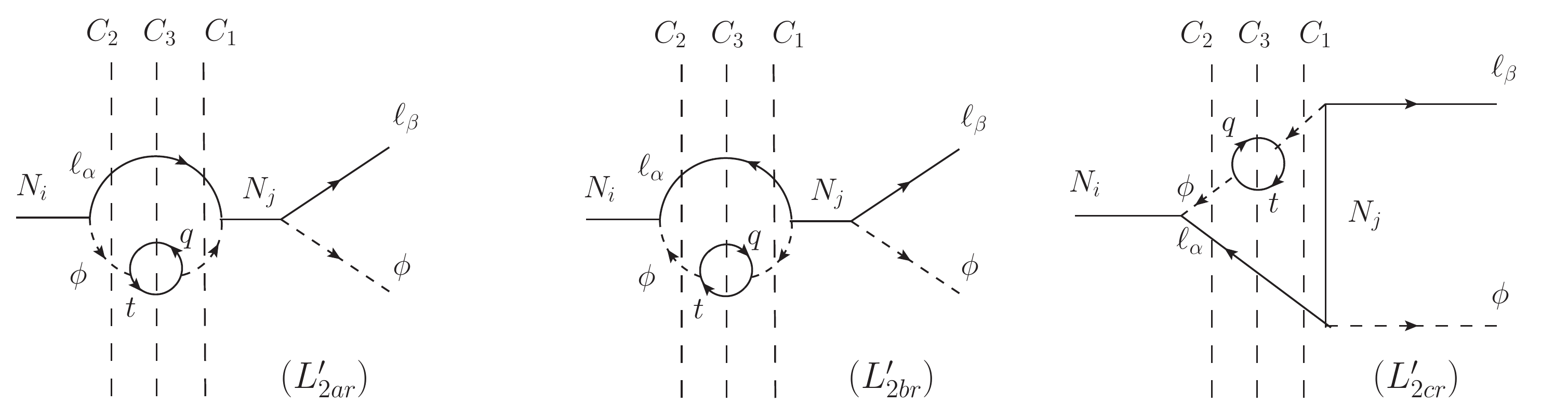,width=0.95\textwidth,angle=0}}}
\caption[]{An additional set of diagrams contributing to the CP asymmetry in two-body decays. The notation follows the conventions explained in the caption of Fig.~\ref{fig:3bd}.} 
\label{fig:2bd}
\end{figure}

The expression for this asymmetry will be given below, but before it is instructive to consider schematically how the cancellations of CP asymmetries and infrared divergences arise. Indeed, the integration of $\abss{A(\proname{N_i}{\ell_\alpha \bar q t})}$ over the phase space of the final massless particles has a collinear divergence when $\bar q$ and $t$ are emitted in the same direction. For massless quarks this final state cannot be distinguished from the corresponding one in the two-body decay  $\proname{N_i}{\ell_\alpha \phi}$, therefore at $\order{\lambda_t^2}$ it is more appropriate to calculate the joint sum of two- and three-body decays. It is precisely this sum the necessary one to cancel infrared divergences following the KLN theorem~\footnote{An insightful analysis about infrared divergences and the KLN theorem has been presented recently in~\cite{Frye:2018xjj}. In particular, the cancellation of infrared divergences in the joint sum of two- and three-body decays of a heavy neutral particle, was related to the unitarity requirement of probabilities adding up to a finite value (one!), and the fact that the forward scattering amplitude for a massive neutral particle is infrared finite. This work also shows a systematic way of finding sets of diagrams that added together yield infrared finite cross sections.} (note that at $\order{\lambda_t^2}$ the quark loop introduces an infrared divergence in $\proname{N_i}{\ell_\alpha \phi}$). An explicit verification was done in~\cite{Salvio:2011sf}  for the computation of NLO corrections to the interaction rates of the singlet neutrinos. In this work we have verified that, as expected, the infrared divergences also cancel in the corresponding sum of CP asymmetries. For instance, 
\begin{equation*}
T_3 L_{3a} + T_2 L_{2ar} + L_{2r} L_{2a} \quad \to \quad {\rm infrared \;  finite.}
\end{equation*}
Here and in the following paragraphs we use a loose notation whereby the same symbols are used to denote the diagrams in Figs.~\ref{fig:3bd},~\ref{fig:2bd} and their corresponding amplitudes, integration over phase space is omitted, and the product refers to an interference term (with complex conjugation and factors of 2  also dropped). The diagrams $L_{2x}$ ($x=a,b,c$) are defined in the caption of Fig.~\ref{fig:3bd}.

Next consider the unitarity requirement $\sum_f  \abss{A(\proname{N_i}{f})} = \sum_f \abss{A(\proname{\bar N_i}{\bar f})}$. Actually this relation is trivial for a Majorana neutrino because $\bar N_i = N_i$ and the sum runs over the same set of final states~\footnote{It is also worth noticing that $\sum_f \Delta \abss{A(\proname{N_i}{f})}=0$ of course does not imply a null source term for the lepton asymmetry (not even after summing over lepton flavours), because, again, the sum runs over all possible final states, that in this case involve antileptons as well as leptons, which enter the BE with opposite signs.}. Still it is interesting to consider how the cancellations arise in pairs in the related identity $\sum_f \Delta \abss{A(\proname{N_i}{f})}=0$, as explained in the previous section. For example, the contribution from the interference of diagram $T_3$ with $L_{3a}$ in Fig.~\ref{fig:3bd}, cancels with the interference of diagrams $T_{2\beta}$ and $L_{2ar}^{\prime \, C_3}$ in Fig.~\ref{fig:2bd}. Here we have introduced two more pieces of notation: an additional subscript $\beta$ in $T_2$ to indicate that the final lepton is of flavour $\beta$ instead of $\alpha$, and a superscript $C_i$ to specify the cut for diagrams with more than one. The complete list of cancelling pairs is
\begin{eqnarray*}
T_3 L_{3a} + T_{2\beta} L^{\prime \, C_3}_{2ar} &= 0 ,\; T_3 L_{3x} + {\bar T}_{2\beta} {\bar L}^{\prime \, C_3}_{2xr} &= 0 \;, \\
T_2 L_{2ar} + T_{2\beta} L^{\prime \, C_2}_{2ar} &= 0 ,\; T_2 L_{2xr} + {\bar T}_{2\beta} {\bar L}^{\prime \, C_2}_{2xr} &= 0 \;,\\
L_{2r} L_{2a} + T_{2\beta} L^{\prime \, C_1}_{2ar} &= 0, \; L_{2r} L_{2x} + {\bar T}_{2\beta} {\bar L}^{\prime \, C_1}_{2xr} &= 0 \;, 
\end{eqnarray*}
with $x=b,c$, and the bar over $T$ or $L$ denotes the CP-conjugate process.

The full computation of the CP asymmetries in the decay rates at $\order{\lambda_t^2}$ are given next, separated into a wave L-conserving piece (subscript ``$\rm{w}, L$'' below, coming from the diagrams with an ``$a$'' subscript in Fig.~\ref{fig:3bd}, with L denoting lepton number), a wave L-violating piece (subscript ``$\rm{w}, \mislash{L}$'' below, coming from the diagrams with a ``$b$'' subscript in Fig.~\ref{fig:3bd}), and a vertex contribution (subscript ``${\rm v}$'' below, coming from the diagrams with a ``$c$'' subscript in Fig.~\ref{fig:3bd}). We have used the $\overline{\rm MS}$ renormalization scheme with the scale $\bar \mu= M_i$ to deal with the ultraviolet divergences, while the infrared ones have been regularized via a small top mass, $m_q = m_t \equiv m$. Moreover, it has been assumed that $\abs{M_i - M_j} \gg \Gamma_{i,j}$, where $\Gamma_k$ denotes the total decay width of $N_k$. Then,

\begin{equation}
\dgmhor{N_i}{\ell_\alpha \bar q t} = \Delta \Gamma^{\rm 3b}_{\rm{w}, L} + \Delta \Gamma^{\rm 3b}_{\rm{w}, \mislash{L}} + \Delta \Gamma^{\rm 3b}_{\rm v} \; ,
\end{equation}
with
\begin{eqnarray}
 \Delta \Gamma^{\rm 3b}_{\rm{w}, L} & = &  \sum_{j \neq i} \lambda_t^2 \; \miim{(\lambda^\dagger \lambda)_{j i} \lambda^*_{\alpha i} \lambda_{\alpha j}} \frac{M_i}{(2 \pi)^4 \, 2^6} \frac{M_i^2}{M_j^2 - M_i^2} \left[- \frac{23}{2} + 6 \ln \frac{M_i}{m}\right] \; , \label{eq:d3bda} \\
 \Delta \Gamma^{\rm 3b}_{\rm{w}, \mislash{L}} & = &  \sum_{j \neq i} \lambda_t^2 \; \miim{(\lambda^\dagger \lambda)_{i j} \lambda^*_{\alpha i} \lambda_{\alpha j}} \frac{M_i}{(2 \pi)^4 \, 2^6} \frac{M_i \, M_j}{M_j^2 - M_i^2} \left[- \frac{23}{2} + 6 \ln \frac{M_i}{m}\right] \; , \label{eq:d3bdb} 
 \end{eqnarray}
 \begin{eqnarray}
 \Delta \Gamma^{\rm 3b}_{\rm v} & = &  \sum_{j \neq i} \lambda_t^2 \; \miim{(\lambda^\dagger \lambda)_{i j} \lambda^*_{\alpha i} \lambda_{\alpha j}} \frac{M_j}{(2 \pi)^4 \, 2^6}  \left[ 13 - 4  f_{i j} +6 (f_{ij} - 1) \ln \frac{M_i}{m} + 3 g_{ij} \right] \notag \\
 &= & \sum_{j \neq i} \lambda_t^2 \; \miim{(\lambda^\dagger \lambda)_{i j} \lambda^*_{\alpha i} \lambda_{\alpha j}} \frac{M_i}{(2 \pi)^4 \, 2^7} \frac{M_i}{M_j} \left[- \frac{23}{2} + 6 \ln \frac{M_i}{m}\right] + \order{\tfrac{M_i^4}{M_j^4}} \; , \label{eq:d3bdc} 
\end{eqnarray}
where
\begin{equation}
f_{i j} = \left( 1 + \frac{M_j^2}{M_i^2}\right) \ln \left( 1 + \frac{M_i^2}{M_j^2}\right) \; ,
\end{equation}
and
\begin{equation}
g_{i j} = g_{i j} (m) = \int_{4 m^2}^{M_i^2} \dif x \; \frac{\sqrt{x - 4 m^2}}{x^{3/2}} \left(1 + \frac{M_j^2}{M_i^2} \right) \ln \left( \frac{M_i^2 + M_j^2 - x}{M_j^2} \right) - 2 f_{i j} \ln \frac{M_i}{m} \; .
\end{equation}
This integral has a finite limit when $m \to 0$, but we could not find an analytical expression.
We have summed over all final degrees of freedom and averaged over the initial ones.

In the last line of Eq.~\ref{eq:d3bdc} we have expanded the expression in powers of $M_i/M_j$, keeping only the lowest order terms, to reveal the hierarchical limit of sterile neutrino masses.

Similarly,
\begin{equation}
\dgmhor{N_i}{\ell_\alpha \phi} = \Delta \Gamma^{\rm 2b}_{\rm{w}, L} + \Delta \Gamma^{\rm 2b}_{\rm{w}, \mislash{L}} + \Delta \Gamma^{\rm 2b}_{\rm v} \; ,
\end{equation}
with
\begin{eqnarray}
 \Delta \Gamma^{\rm 2b}_{\rm{w}, L} & = &  \sum_{j \neq i} \miim{(\lambda^\dagger \lambda)_{j i} \lambda^*_{\alpha i} \lambda_{\alpha j}} \frac{M_i}{(2 \pi)^2 \, 2^4} \frac{M_i^2}{M_j^2 - M_i^2} \left[1+ \frac{\lambda_t^2}{(2\pi)^2 2^2} \left(1 - 6 \ln \frac{M_i}{m}\right) \right], \label{eq:d2bda} \\
 \Delta \Gamma^{\rm 2b}_{\rm{w}, \mislash{L}} & = &   \sum_{j \neq i} \miim{(\lambda^\dagger \lambda)_{i j} \lambda^*_{\alpha i} \lambda_{\alpha j}} \frac{M_i}{(2 \pi)^2 \, 2^4} \frac{M_i M_j}{M_j^2 - M_i^2} \left[1+ \frac{\lambda_t^2}{(2\pi)^2 2^2} \left(1 - 6 \ln \frac{M_i}{m}\right) \right], \label{eq:d2bdb} \\
 \Delta \Gamma^{\rm 2b}_{\rm v} & = &   \sum_{j \neq i} \miim{(\lambda^\dagger \lambda)_{i j} \lambda^*_{\alpha i} \lambda_{\alpha j}} \frac{M_j}{(2 \pi)^2 \, 2^4} (f_{i j} - 1) \left[1+ \frac{\lambda_t^2}{(2\pi)^2 2^2} \left(1 - 6 \ln \frac{M_i}{m}\right) \right], \label{eq:d2bdc} \\
 &= & \sum_{j \neq i} \miim{(\lambda^\dagger \lambda)_{i j} \lambda^*_{\alpha i} \lambda_{\alpha j}} \frac{M_i}{(2 \pi)^2 \, 2^5} \frac{M_i}{M_j} \left[1+ \frac{\lambda_t^2}{(2\pi)^2 2^2} \left(1 - 6 \ln \frac{M_i}{m}\right) \right] + \order{\tfrac{M_i^4}{M_j^4}}. \notag
\end{eqnarray}
It is immediate to verify from these expressions that the infrared divergences cancel when adding corresponding contributions to the two- and three-body decay asymmetries.
 
The CP asymmetries, $\epsilon \equiv \tfrac{\gmhor{i}{j} - \gmhor{\bar i}{\bar j}}{\sum_j \gmhor{i}{j} + \gmhor{\bar i}{\bar j}}$, can be obtained dividing the expressions above by the decay widths,
\begin{eqnarray}
\Gamma^{\rm 2b}  =  \sum_\alpha \gmhor{N_i}{\ell_\alpha \phi} + \gmhor{N_i}{\bar \ell_\alpha \bar \phi} &=& \frac{(\lambda^\dagger \lambda)_{i i}}{8 \pi} M_i \left[1+ \frac{\lambda_t^2}{(2\pi)^2 2^2} \left(1 - 6 \ln \frac{M_i}{m}\right) \right]\, , \notag \\
\Gamma^{3b}  =  \sum_\alpha \gmhor{N_i}{\ell_\alpha \bar q t} + \gmhor{N_i}{\bar \ell_\alpha q \bar t} &=&   \frac{(\lambda^\dagger \lambda)_{i i}}{8 \pi} M_i\frac{\lambda_t^2}{(2\pi)^2 2^2} \left[-\frac{23}{2} + 6 \ln \frac{M_i}{m} \right]\,.
\end{eqnarray}

\begin{figure}[!t]
\centerline{\protect\hbox{
\epsfig{file=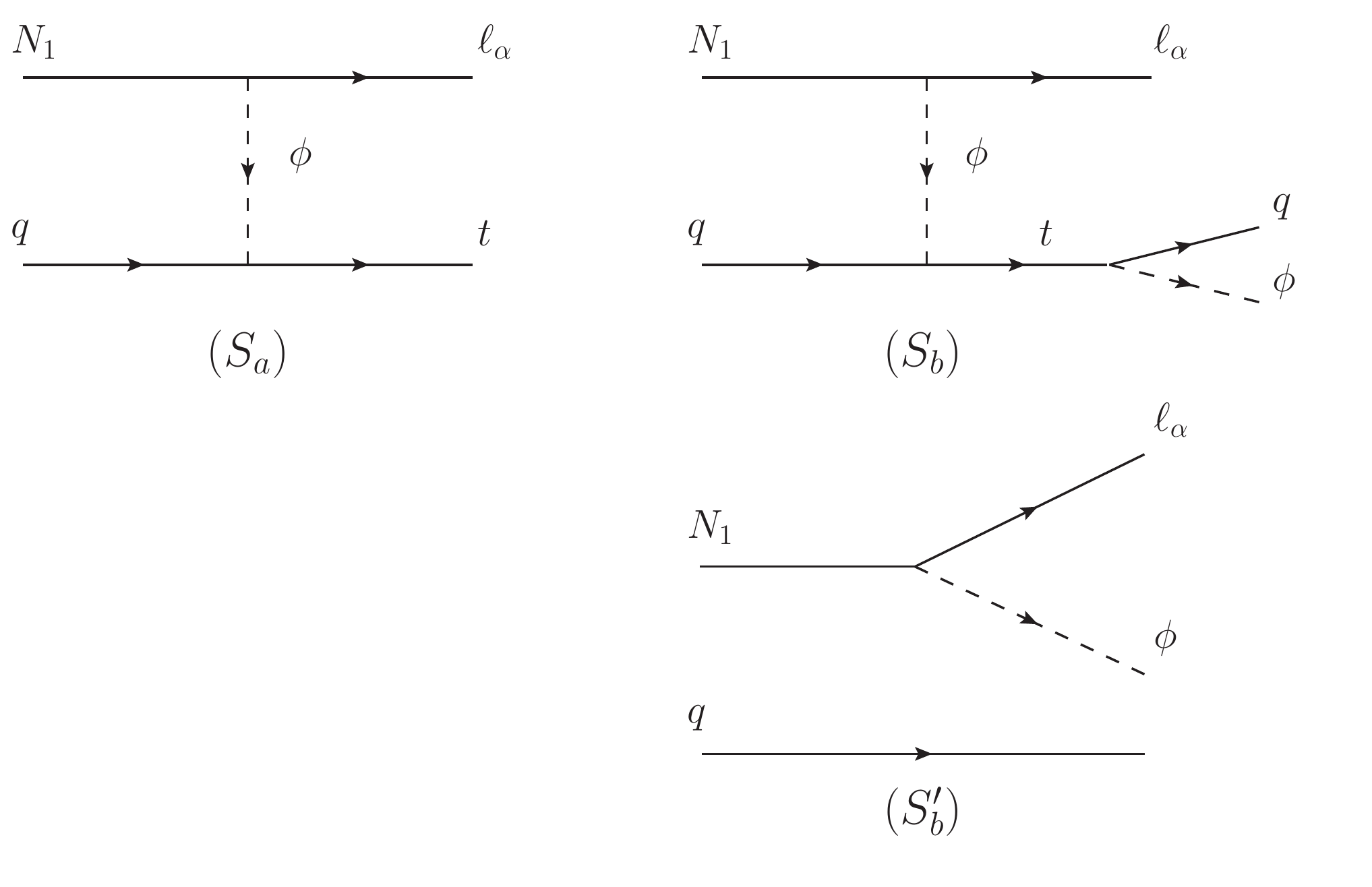,width=0.70\textwidth,angle=0}}}
\caption[]{Feynman diagrams for the scatterings $\proname{N_1 q}{\ell_\alpha t}$ and $\proname{N_1 q}{\ell_\alpha q \phi}$,  which should be considered concomitantly in order to cancel infrared divergences. At $\order{\lambda_t^2}$ the process $\proname{N_1 q}{\ell_\alpha q \phi}$ receives a contribution from the interference of diagrams $S_b$ and $S_b^\prime$.} 
\label{fig:irdiv}
\end{figure}

In order to obtain the complete source term in the BE at $\order{\lambda_t^2}$, it is necessary to consider many more CP-violating processes, being especially careful to include the contributions from disconnected diagrams. This is out of the goal of this paper, but we wish to make a final comment. Several of these processes have infrared divergences. This is, e.g., the case with the scattering $\proname{N_1 q}{\ell_\alpha t}$. For massless quarks there is a collinear divergence when the momenta of $q$ and $t$ are parallel. This is related to the fact that a massless $t$ quark cannot be distinguished from a pair of massless $\phi$ and $q$ with the same momenta. Therefore, following the KLN theorem, it is more appropriate to consider the processes $\proname{N_1 q}{\ell_\alpha t}$ and $\proname{N_1 q}{\ell_\alpha q \phi}$ together. At $\order{\lambda_t^2}$, the process $\proname{N_1 q}{\ell_\alpha q \phi}$ gets a contribution from the interference of the diagrams depicted in Fig.~\ref{fig:irdiv}. The integral over phase space must be handled with care, because the disconnected quark line enforces the intermediate Higgs propagator in diagram $S_b$ to be on-shell. A neat way to deal with this issue has been explained very recently in~\cite{Frye:2018xjj}. Following the prescription in the Appendix A of~\cite{Frye:2018xjj}, we have verified that indeed the infrared divergences cancel when summing the rates of both processes in Fig.~\ref{fig:irdiv}. Namely, if we denote by $P, Q, p, r, Q',$ and $k$ the 4-momentum of the external $N_1, q$ (initial), $\ell_\alpha, t, q$ (final), and $\phi$, respectively, and by $s$ the squared center-of-mass energy, then the integrals over final-state phase space of the corresponding contributions to the squared amplitudes, summing over the spin of initial and final particles, are equal to, in the center-of-mass frame,
\begin{eqnarray}
I_a & \equiv & \sum_{{\rm spin}}  \int  \dif \pi_{\ell_\alpha} \dif \pi_t \, (2 \pi)^4  \delta^4(p + r - P - Q) \abss{A(S_a)} = \frac{1}{16 \pi} \frac{\lambda_t^2 \abss{\lambda_{\alpha 1}}}{\sqrt{s} \abs{\bf P}} M_1^2 \ln \frac{M_1^2}{m^2} + {\rm IRF} \, ,\notag \\ & & \\
I_b & \equiv & \sum_{{\rm spin}}  \int \dif \pi_{\ell_\alpha} \dif \pi_\phi \dif \pi_q \, (2 \pi)^4 \delta^4(p + k + Q' - P - Q) \left[ A(S_b) A^*(S_b^\prime) + {\rm c.c.} \right] =  \notag \\ 
& &\frac{1}{4 \pi} \frac{\lambda_t^2 \abss{\lambda_{\alpha 1}}}{\abs{\bf P}} \int \dif \!\abs{\bf k} \dif k^0 \, \delta(k^2) \, \theta(k^0) \left( \frac{1}{k^2 + i \epsilon} + {\rm c.c.}\right) \, f(k^0) \, ,
\end{eqnarray}   
where
\begin{equation*}
f(k^0) = \abs{\bf k} \left(2 M_1^2 - 2kP \right) \frac{2kQ + 2m^2}{2kQ + k^2} \, .
\end{equation*}
The 4-momentum of $\phi$ has been written as $k=(k^0,{\bf k})$, so that, e.g., $k^2 = k^{0\, 2} - {\bf k}^2 \equiv k^{0\, 2} - \abss{\bf k}$, and analogously for the 4-momenta of the other particles. Moreover, infrared-finite terms, which are not relevant to the present discussion, have been denoted by IRF. As shown in~\cite{Frye:2018xjj}, the action of the distribution $\delta(k^2) \, \theta(k^0) \left( \tfrac{1}{k^2 + i \epsilon} + {\rm c.c.}\right)$ on the test function $f(k^0)$ is a derivative. More precisely,
\begin{eqnarray}
I_b & = & \frac{1}{4 \pi} \frac{\lambda_t^2 \abss{\lambda_{\alpha 1}}}{\abs{\bf P}} \int \dif\!\abs{\bf k} \, \frac{\dif}{\dif k^0} \left[ \frac{f(k^0)}{(k^0 + \abs{\bf k})^2}\right]_{k^0=\abs{\bf k}} \notag \\
& = & -\frac{1}{16 \pi} \frac{\lambda_t^2 \abss{\lambda_{\alpha 1}}}{\sqrt{s} \abs{\bf P}} M_1^2 \ln \frac{M_1^2}{m^2} + {\rm IRF} \;.
\end{eqnarray}
Hence, the infrared-divergent terms cancel in the sum $I_a+I_b$.

As a final remark, notice that the CP asymmetry in the process $\proname{N_1 q}{\ell_\alpha q \phi}$, not included in previous works, is a priori of similar size than the CP asymmetry in $\proname{N_1 q}{\ell_\alpha t}$, due again to the interference of connected with disconnected diagrams.  

\section{Conclusions and outlook}
\label{sec:con}

Using unitarity and CPT invariance we have derived in a simple way the source term of the BE for the density asymmetry of some particle denoted generically by $a$ (e.g. $a$ can be a SM lepton doublet in leptogenesis). Processes that do not change neither the number nor the momenta of $a$, like $Y a(p) \to X a(p)$, must be considered with care, given that they can yield finite contributions to the integrals over phase space due to the interference of connected with disconnected diagrams. In order to apply the unitary conditions, it may be convenient to include them twice in the BE for $f_a$, once with a negative sign (corresponding to the destruction of an $a(p)$), and another with a positive sign (corresponding to the production of an $a(p)$). Under this convention, we showed that the only contributions to the source of the linearized BE for $\Delta f_a$ come from production processes of ``$a$'' (or ``$\bar a$'') particles, with an out-of-equilibrium species in the initial state. We also discussed on the number of different out-of-equilibrium species necessary to generate a baryon asymmetry from scatterings, following an argument started in~\cite{baldes14}.  In Sec.~\ref{sec:cut} we showed how to easily obtain pairs of cancelling contributions to the CP asymmetries. This allowed us to find new CP-violating processes at first order in the top Yukawa coupling, but also important cancellations. Notably, some of these involve the interference of connected with disconnected diagrams.  In Sec.~\ref{sec:3bd} we calculated the CP asymmetry in the three-body decay $\proname{N}{\ell \bar q t}$, considering the processes that should be simultaneously included to cancel infrared divergences as required by the KLN theorem, and providing detailed examples of the issues discussed in previous sections. 

It has been out of the goal of this paper to compute the full source term at first order in the top Yukawa coupling, i.e at $\order{\lambda_t^2}$.  At low temperatures compared to the mass of the lightest sterile neutrino, $T \lesssim M$, the quantitative effect is expected to be small, of order a few $\%$ (check, e.g., the studies performed in several of the papers cited in the introduction). Nevertheless, we think one interesting reason to make such a complete calculation would be to check the results of the novel approach in~\cite{Bodeker:2017deo}, where a relation between the CP-violating rates and finite-temperature real-time correlation functions was derived, and explicit expressions in the hierarchical limit of sterile neutrino masses were obtained. The analysis in our work can also be useful to calculate the full source term at first order in the SM gauge couplings, completing the results of~\cite{Fong:2010bh}.

\section*{Acknowledgments}
We wish to thank Esteban Roulet for useful comments and the referee for constructive suggestions.

\bibliographystyle{JHEP}
\bibliography{referencias_leptogenesis3}

\end{document}